# Cosmology: small scale issues

Joel R. Primack[a]

[a]*Physics Department, University of California, Santa Cruz, CA 95064 USA*

**Abstract.** The abundance of dark matter satellites and subhalos, the existence of density cusps at the centers of dark matter halos, and problems producing realistic disk galaxies in simulations are issues that have raised concerns about the viability of the standard cold dark matter (ΛCDM) scenario for galaxy formation. This talk reviews these issues, and considers the implications for cold vs. various varieties of warm dark matter (WDM). The current evidence appears to be consistent with standard ΛCDM, although improving data may point toward a rather tepid version of ΛWDM – tepid since the dark matter cannot be very warm without violating observational constraints.

## DARK MATTER IS OUR FRIEND

Dark matter preserved the primordial fluctuations in cosmological density on galaxy scales that were wiped out in baryonic matter by momentum transport (viscosity) as radiation decoupled from baryons in the first few hundred thousand years after the big bang. The growth of dark matter halos started early enough to result in the formation of galaxies that we see even at high redshifts $z > 6$. Dark matter halos provide most of the gravitation within which stable structures formed in the universe. Dark matter halos preserve these galaxies, groups, and clusters as the dark energy tears apart unbound structures and expands the space between bound structures such as the Local Group of galaxies. Thus we owe our existence and future to dark matter.

Cold dark matter theory [1] including cosmic inflation has become the basis for the standard modern ΛCDM cosmology, which is favored by analysis of the available cosmic microwave background data and large scale structure data over even more complicated variant theories having additional parameters [2]. Most of the cosmological density is nonbaryonic dark matter (about 23%) and dark energy (about 72%), with baryonic matter making up only about 4.6% and the visible baryons only about 0.5% of the cosmic density. The fact that dark energy and dark matter are dominant suggests a popular name for the modern standard cosmology: the "double dark" theory, as Nancy Abrams and I suggested in our recent book about modern cosmology and its broader implications [3].

The physical nature of dark matter remains to be discovered. The two most popular ideas concerning the identity of the dark matter particles remain the lightest supersymmetric partner particle [4], also called supersymmetric weakly interacting massive particles (WIMPs) [5], and the cosmological axion [6], recently reviewed in [7]. These are the two dark matter candidate particles that are best motivated in the sense that they are favored by other considerations of elementary particle theory.

Supersymmetry remains the best idea for going beyond the standard model of particle physics. It allows control of vacuum energy and of otherwise unrenormalizable gravitational interactions, and thus may allow gravity to be combined with the electroweak and strong interactions in superstring theory. Supersymmetry also allows for grand unification of the electroweak and strong interactions, and naturally explains how the electroweak scale could be so much smaller than the grand unification or Planck scales (thus solving the "gauge hierarchy problem"). It thus leads to the expectation that the supersymmetric WIMP mass will be in the range of about 100 to 1000 GeV.

Axions remain the best solution to the CP problem of SU(3) gauge theory of strong interactions, although it is possible that the axion exists and solves the strong CP problem but makes only a negligible contribution to the dark matter density.

Many other particles have been proposed as possible dark matter candidates, even within the context of supersymmetry. An exciting prospect in the next few years is that experimental and astronomical data may point toward specific properties of the dark matter particles, and may even enable us to discover their identity. Other talks at this conference discussed direct detection experiments, typically involving detecting elastic scattering of WIMPs by nucleons or detecting the conversion of axions to photons in a strong magnetic field, and indirect detection experiments such as detecting gamma rays or positrons produced by WIMP annihilation. The present lecture is concerned with potential problems for CDM and clues to the nature of the dark matter from astronomical data such as substructure within dark matter halos, especially subhalos and satellites, central cusps, and angular momentum issues.

## SUBHALOS AND SATELLITES

It at first seemed plausible that the observed bright satellite galaxies are hosted by the most massive subhalos of the dark matter halo of the central galaxy, but this turned out to predict too large a radial distribution for the satellite galaxies. Andrey Kravtsov and collaborators [8] proposed instead that bright satellite galaxies are hosted by the subhalos that were the most massive when they were accreted. This hypothesis appears to correctly predict the observed radial distribution of satellite galaxies, and also of galaxies within clusters. It also explains naturally why nearby satellites are dwarf spheroidals (dSph) while more distant ones are a mix of dwarf spheroidal and dwarf irregular galaxies [8].

An issue that is still regularly mentioned by observational astronomers (e.g. [9]) as a problem for ΛCDM is the fact that many fewer satellite galaxies have been detected in the Local Group than the number of subhalos predicted. But developing theory and the recent discovery of many additional satellite galaxies around the Milky Way and the Andromeda galaxy suggest that this is not a problem at all (e.g. [10]). As Figure 1(a) shows, it is only below a circular velocity ~30 km s$^{-1}$ that the number of dark matter halos begins to exceed the number of observed satellites. Figure 1(b) shows that suppression of star formation in small dwarf galaxies after reionization can account for the observed satellite abundance in ΛCDM, as suggested by [11-14]. Whether better understanding of such baryonic physics can also explain the recent discovery [15] that all the local faint satellites have roughly the same dynamical mass of about $10^7$ solar masses within their central 300 parsecs remains to be seen. Alternatively, it is possible that this reflects a clustering scale in the dark matter, which would be a clue to its nature. The newly discovered dwarf satellite galaxy properties such as metallicity appear continue the scaling relations discovered earlier, with metallicity decreasing with luminosity [18]. Explaining this is challenge [19] for theories of the formation of satellite galaxies.

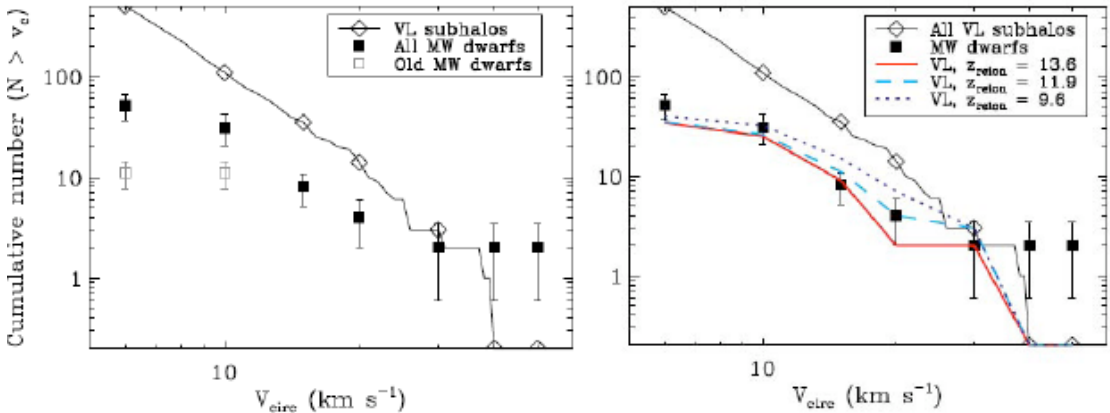

**FIGURE 1.** (a) Cumulative number of Milky Way satellite galaxies as a function of halo circular velocity, assuming Poisson errors on the number count of satellites in each bin. The filled black squares include the new circular velocity estimates from [10], who follow [16] and use $V_{circ} = \sqrt{3}\,\sigma$. Diamonds represent all subhalos within the virial radius in the Via Lactea simulation [17]. (b) Effect of reionization on the missing satellite problem. The lower solid curve shows the circular velocity distribution for the 51 most massive Via Lactea subhalos if reionization occurred at $z = 13.6$, the dashed curve at $z = 11.9$, and the dotted curve at $z = 9.6$. (Figures from [10].)

Hogan and Dalcanton [20] introduced the parameter $Q = \rho/\sigma^3$ as an estimate of the coarse-grained phase-space density of the dark matter in galaxy halos. Liouville's theorem implies that observed values of Q set a hard lower limit on the original phase-space density of the dark matter. All of the galaxies except UMa I, CVn I, and Hercules have $Q > 10^{-3}$ $M_{sun}$ $pc^{-3}$ $(km\ s^{-1})^{-3}$, about an order of magnitude improvement compared to the previously-known dSphs. The subhalos in Via Lactea II [21] that could host Milky Way satellites have densities and phase space densities comparable to these values. This places significant limits on non-CDM dark matter models; for example, it implies that the mass of a WDM particle must be $m_x > 1.2$ keV.

Via Lactea II is the highest resolution simulation of a galaxy-mass halo published at this writing, and it is able to resolve substructure even at the distance of the sun from the center of the Milky Way. The fraction of mass in the Via Lactea II subhalos of mass $\sim 10^6 - 10^8$ $M_{sun}$ is ~0.5%. This is about the amount needed to explain the flux anomalies observed in radio images of quasars that are quadruply gravitationally lensed by foreground elliptical galaxies [22]. Free streaming of WDM particles can considerably dampen the matter power spectrum in this mass range, so a WDM model with an insufficiently massive particle (e.g., a standard sterile neutrino $m_v < 10$ keV) fails to reproduce the observed flux anomalies [23]. In order to see whether this is indeed a serious constraint for WDM and a triumph for CDM, we need more than the 6 radio quads now known – a challenge for radio astronomers!

An additional constraint on WDM comes from reionization. While the first stars can reionize the universe starting at redshift $z > 20$ in standard $\Lambda$CDM [24], the absence of low mass halos in $\Lambda$WDM delays reionization [25]. Reionization is delayed significantly in $\Lambda$WDM even with $m_x = 15$ keV [26]. The actual constraint on $m_x$ from the cosmic microwave background and other data remains to be determined. If the WDM is produced by decay of a higher-mass particle, the velocity distribution and phase space constraints can be different [27,28]. MeV dark matter, motivated by observation of 511 keV emission from the galactic bulge, also can suppress formation of structure with masses up to about $10^7$ $M_{sun}$ since such particles are expected to remain in equilibrium with the cosmic neutrino background until relatively late times [29].

Sterile neutrinos that mix with active neutrinos are produced in the early universe and could be the dark matter [30]. Such neutrinos would decay into X-rays plus light neutrinos, so non-observation of X-rays from various sources gives upper limits on the mass of such sterile neutrinos $m_s < 3.5$ keV. Since this upper limit is inconsistent with the lower limit $m_s > 28$ keV from Lyman-alpha-forest data [31], that rules out such sterile neutrinos as the dark matter, although other varieties of sterile neutrinos are still allowed and might explain neutron star kicks [32,33].

Note finally that various authors [34-36] have claimed that $\Lambda$WDM substructure develops in simulations on scales below the free-streaming cutoff. If true, this could alleviate the conflict between the many small subhalos needed to give the observed number of Local Group satellite galaxies, taking into account reionization and feedback, and needed to explain gravitational lensing radio flux anomalies. However Wang and White [37] recently showed that such substructure arises from discreteness in the initial particle distribution, and is therefore spurious.

As a result of the new constraints just mentioned, it follows that the hottest varieties of warm dark matter are now ruled out, so if the dark matter is not cold (i.e., with cosmologically negligible constraints from free-streaming, as discussed in the original papers that introduced the hot-warm-cold terminology [1,38]) then it must at least be rather tepid.

## CUSPS IN GALAXY CENTERS

This was first recognized as a potential problem for CDM by Flores and me [39] and by Moore [40]. However, beam smearing in radio observations of neutral hydrogen in galaxy centers was significantly underestimated [41,42] in the early observational papers; taking this into account, the observations imply an inner density $\rho \propto r^{-\alpha}$ with slope satisfying $0 \leq \alpha < 1.5$, and thus consistent with the $\Lambda$CDM Navarro-Frenk-White [43] slope $\alpha$ approaching 1 from above at small radius $r$. Low surface brightness galaxies are mainly dark matter, so complications of baryonic physics are minimized but could still be important [44,45]. A careful study of the kinematics of five nearby low-mass spiral galaxies found that four of them had significant non-circular motions in their central regions; the only one that did not was consistent with $\alpha \approx 1$ [46] as ΛCDM predicts. The central non-circular motions observed in this galaxy sample and many others could be caused by nonspherical halos [47]. Dark matter halos are increasingly aspherical at smaller radii, at higher redshift, and at larger masses [48-51] and this can perhaps account for the observed rotation curves [52-54]. However, only fully self-consistent ΛCDM simulations of small spiral galaxies including all relevant baryonic physics, which can modify the central density distributions and thus the kinematics, will be able to tell whether they are fully consistent with observations.

Observations indicate that dark matter halos may also be too concentrated farther from their centers [55]. Halos hosting low surface brightness galaxies may have higher spin and lower concentration than average [56,51], which would improve agreement between ΛCDM predictions and observations. It remains unclear how much adiabatic contraction [57-59] occurs as the baryons cool and condense toward the center, since there are potentially offsetting effects from dynamical friction and blowout of gas. Recent analyses comparing spiral galaxy data to theory conclude that there is little room for adiabatic contraction [60,61], and that a bit of halo expansion may better fit the data [61]. Early ΛCDM simulations with high values $\sigma_8 \sim 1$ of the linear mass fluctuation amplitude in spheres of 8 $h^{-1}$ Mpc (a measure of the amplitude of the power spectrum of density fluctuations) predicted high concentrations [62], which are lower with lower values of $\sigma_8$ [63]. The cosmological parameters from WMAP5 and large scale structure observations [2], including $\sigma_8 = 0.82$, lead to concentrations that match galaxy observations better [64], and they may also match observed cluster concentrations [65,66].

## ANGULAR MOMENTUM ISSUES

The growth of the mass of dark matter halos and its relation to the structure of the halos has been studied based on structural merger trees [56], and the angular momentum of dark matter halos is now understood to arise largely from the orbital angular momentum of merging progenitor halos [67,68]. But it is now clear that the baryonic matter in disk galaxies has an angular momentum distribution very different from that of the dark matter [69,70]. Although until recently simulations were not able to account for the formation and structure of disk galaxies, simulations with higher resolution and improved treatment of stellar feedback from supernovae are starting to produce disk galaxies that resemble those that nature produces [71,72]. It remains to be understood how the gas that forms stars acquires the needed angular momentum. Possibly important is the recent realization that gas enters lower-mass halos cold and in clouds [73,74], rather than being heated to the halo virial temperature as in the standard treatment used in semi-analytic models [1,75].

# SMALL SCALE ISSUES: SUMMARY

**Satellites**: The discovery of many faint Local Group dwarf galaxies is consistent with ΛCDM predictions. Reionization, lensing, satellites, and Lyman-alpha forest data imply that if the dark matter is WDM, it must be tepid at most – i.e., not too warm.

**Cusps:** The triaxial nature of dark matter halos plus observational biases suggests that the observed velocity structure of low surface brightness and dwarf spiral galaxies are consistent with cuspy ΛCDM halos.

**Angular Momentum:** ΛCDM simulations are increasingly able to form realistic spiral galaxies, as resolution improves and feedback becomes more realistic.

# ACKNOWLEDGMENTS

I thank NASA and NSF for grants that supported research relevant to this topic, and I thank DOE and NASA for supercomputer allocations for relevant simulations. My group's simulations were also done on the Pleiades cluster at UCSC, supported by an NSF MRI grant. I thank my students and other collaborators for many helpful discussions, and David Cline for organizing the very useful Sources of Dark Matter conferences and inviting me to give this talk.